\documentclass[10pt]{article}
\usepackage{jheppub2}  
\pdfoutput=1

\usepackage{graphicx}     
\usepackage{multirow}
\usepackage{booktabs}
\usepackage{float}        
\usepackage{subcaption}   
\usepackage{caption}      
\usepackage[utf8]{inputenc} 
\usepackage{hyperref}     
\usepackage{amsmath, amssymb, amsthm} 

\usepackage{xcolor}
\definecolor{cherryblossompink}{rgb}{1.0, 0.72, 0.77}
\definecolor{lightblue}{rgb}{0.68, 0.85, 0.9}

\usepackage{esint}
\usepackage{mathrsfs}
\usepackage{mathtools}
\usepackage{amsfonts}
\usepackage{lmodern}
\usepackage{cancel}
\usepackage{accents}
\usepackage{soul}
\usepackage{comment}
\usepackage[titletoc,title]{appendix}
\setstcolor{red}

\usepackage{tikz}
\usetikzlibrary{positioning}
\usetikzlibrary{intersections}
\usetikzlibrary{fadings} 
\usetikzlibrary{arrows.meta} 
\usetikzlibrary{arrows}
\usetikzlibrary{decorations.pathmorphing}
\usetikzlibrary{decorations.pathreplacing,decorations.markings}
\usetikzlibrary{backgrounds,automata}

\title{Covariant diffusion and drift of the stochastic GW background with LISA}
\author{Giorgio Mentasti,}
\emailAdd{g.mentasti21@imperial.ac.uk}
\author{Arad Nasiri}
\emailAdd{a.nasiri21@imperial.ac.uk}

\date{May, 2025 \textcolor{blue}{To be changed}}
\affiliation{
Blackett Laboratory, Imperial College London, SW7 2AZ, UK}

\abstract{
    We study the covariant diffusion and drift of massless particles on the light cone within the context of quantum gravity phenomenology. Unlike modified dispersion relations that violate Lorentz invariance and grow with frequency, this model introduces a stochastic correction to the massless geodesic equation while preserving Lorentz invariance, and is dominant at lower frequencies due to the larger spacetime support of long-wavelength modes.
    The effect is phenomenologically described by just two diffusion and drift parameters, $\kappa_1$ and $\kappa_2$, whose values are already constrained by measurements of the CMB blackbody spectrum. We show that a direct measurement and characterization of a gravitational wave (GW) background frequency spectrum can improve bounds on these diffusion and drift parameters by over 12 orders of magnitude compared to those from the CMB. In particular, we find that detecting a GW background sourced by realistic models of first-order phase transitions or primordial black holes (PBH) with LISA can constrain the parameters down to a value of $\kappa_1,\,\kappa_2\lesssim  10^{-56}\,\text{kg}\,\text{m}^2\text{s}^{-3}$.
}

\begin{document}

\maketitle

\section{Introduction}

One of the aims of quantum gravity research is to identify the fundamental degrees of freedom underlying spacetime at the Planck scale. Direct experimental probes of such an extreme scale are out of reach. Even with a mathematically satisfactory theory, deriving concrete observable predictions from first principles is difficult. This makes it essential to develop effective, theory-agnostic phenomenological models. These models must satisfy consistency conditions and symmetries while capturing the potential macroscopic signatures of a broad range of quantum gravity theories. Such models can then be directly tested, allowing constraints on their effective parameters. Gravitational waves offer a unique observational window for probing such models.

With the detection of numerous black hole merger events by LIGO and the upcoming plans for the Einstein Telescope and the Laser Interferometer Space Antenna (LISA), there is growing interest in using gravitational waves to test quantum-gravity-inspired models beyond general relativity. While some studies have focused on modifications to the GW sources \cite{kobakhidze2016constraining,maselli2019micro,giddings2019exploring,agullo2021potential}, many have concentrated on the propagation effects of a quantum spacetime on GWs. Although such effects are expected to be small, they can accumulate and amplify over the cosmological distances traversed by GWs. Some studies are mostly motivated by models that arise in the context of a particular quantum gravity theory or modified gravity \cite{garcia2021propagation,calcagni2017lorentz,wang2020exploring}, while others highlight their theory-agnostic approach \cite{yunes2016theoretical,calcagni2019gravitational,calcagni2022quantum,belgacem2019testing}. 

A common theme in many of these studies is modifying the massless dispersion relation and hence effectively violating Lorentz invariance. Current bounds on Lorentz violations of gravitons from LIGO events remain weak \cite{ellis2016remarks}; however, in the case of photons, studies on short gamma-ray bursts \cite{Vasileiou2013vra,Ellis2018lca,LHAASO2024lub,Piran2023xfg} and flares from active galactic nuclei \cite{MAGIC2007etg,HESS2011aa} have placed stringent, in cases trans-Planckian, constraints on the energy scale of such violations. Given these constraints, it is sensible to focus on Lorentz-invariant modifications to graviton propagation. Assuming Lorentz invariance is exact even at the effective level, gravitons must propagate along the light cone but may still experience drift and diffusion due to quantum gravity.

Dowker et. al. introduced a phase space diffusion equation for massless particles in \cite{philpott2009energy}. Although the authors were motivated by considerations in causal set theory, their diffusion model is general, being the unique Lorentz-invariant correction to the Boltzmann equation in flat spacetime. This was further reinforced in \cite{albertini2025stochastic}, where it was shown that the Lorentz-invariant diffusion is equivalent to a stochastic correction to the massless geodesic equation. The uniqueness of the effect, however, is lost in curved spacetime where there can be curvature corrections to the covariance of the noise. Insisting on minimal coupling brings back the uniqueness and makes it the leading order correction to the geodesic equation. Note once again that the massless dispersion equation is preserved. Therefore, any Lorentz-invariant quantum gravity effect on massless particles would be effectively described by this model. In addition, in \cite{philpott2009energy}, the authors used the blackbody spectrum of CMB photons to constrain the model's two free parameters. The most general Lorentz-invariant fluctuations in photon polarization were modeled in \cite{Contaldi2010fh} and constrained using CMB polarization spectra. 

Here we present an analytic study of the massless diffusion equation in FRW spacetime, showing that the strength of the effect is proportional to the inverse of the frequency. This is in stark contrast with the Lorentz-violating effects, which are stronger for large frequencies and are best probed using high-energy gamma rays. However, the variance of the diffusion is proportional to the elapsed affine time. The affine time, defined with a suitable normalization, measures the spacetime area swept by one wavelength \cite{philpott2009energy} and is Lorentz invariant. The larger this area, the greater the stochastic Planck-scale effects on the propagating massless particle. Therefore, for a fixed diffusion constant and propagation distance, lower-frequency waves are the most promising for detecting the diffusion of gravitons. Since LISA probes GW frequencies in the mHz range, we must be able to improve the bounds on the diffusion and drift constants by many orders of magnitude compared to those derived from $\sim100$GHz CMB photons. Indeed, our forecast shows that, in the presence of a GW background with non-trivial spectral features, LISA could tighten the bounds on the diffusion and drift constants by over $\sim$12 orders of magnitude relative to those in \cite{philpott2009energy}.

In principle, one can put bounds on the diffusion effect with LIGO, Virgo, KAGRA, or other ground-based interferometers, since they are already operating and probe frequencies just about 4 orders of magnitude greater than LISA's. However, LIGO has not yet detected a GW background, and there are no universally accepted strong candidates for cosmological sources at those frequencies.
On the other hand, the background sourced by astrophysical events in the LIGO band is not well described by a stochastic ensemble of gravitons but instead arises from the superposition of many unresolved weak coherent events. In addition, resolved sources such as compact binary coalescences, are not describable as a statistical ensemble of GWs. Since a Lorentz invariant diffusion at the field level for $h_{\mu\nu}$ (which is necessary for a resolved source search) is currently lacking, we cannot constrain the diffusion parameter with LIGO catalogs data at this time.

LISA is the first space-based GW interferometer, scheduled for launch in the 2030s. With its unprecedented sensitivity in the millihertz frequency band, LISA will open a new observational window onto the early universe, probing stochastic GW backgrounds of cosmological origin that are inaccessible to ground-based detectors. Cosmological sources such as first-order phase transitions, primordial black hole (PBH) formation, and inflationary processes can generate relic GW backgrounds whose spectral features encode fundamental physics at energy scales far beyond the reach of particle colliders. LISA’s capability to measure the spectral shape, amplitude, and anisotropies of a stochastic gravitational-wave background makes it a unique tool to test early-universe scenarios and constrain new physics models~\cite{Caprini:2015zlo,Bartolo:2016ami,Bartolo:2017p0n}. Strong first-order phase transitions in the early universe arise in many extensions of the Standard Model, notably in scenarios with extended scalar sectors or dark sectors. They serve as a prime example of cosmological signals potentially observable by LISA.  During a first-order phase transition, bubbles of the new vacuum nucleate, expand, and collide, sourcing a stochastic GW signal characterized by a broken power-law spectrum with a peak frequency set by the transition temperature and dynamics of the bubble walls. LISA is particularly sensitive to transitions occurring at energy scales from tens of GeV up to tens of TeV, offering a probe of electroweak-scale and beyond–Standard-Model physics~\cite{Caprini:2015zlo}. As the second example, primordial black holes, hypothesized to form from large density perturbations in the radiation-dominated era, represent another potential source of cosmological GW backgrounds in the LISA band. A population of PBHs generically produces gravitational waves through mechanisms such as Hawking radiation-induced gravitons, PBH mergers, and induced scalar perturbations. The resulting spectra also feature broken power-law shapes, with characteristic frequencies determined by the PBH mass distribution and the underlying primordial power spectrum~\cite{Bartolo:2017p0n}.
These two primordial power spectra candidates, with grounded theoretical motivation, nontrivial frequency features in their spectrum, and sourced in the early universe, can carry a clear footprint of the aforementioned graviton covariant diffusion effect during their propagation.

The paper is organized as follows. In section \ref{SGWB_sec}, we introduce the main quantities used to characterize the stochastic GW background and link them to the graviton distribution function in phase space. In section \ref{Boltzmann_sec}, we review the modified massless geodesic equation and Boltzmann equation in the presence of a Lorentz invariant diffusion effect. In section \ref{analytic_sol_sec}, we analytically solve the modified Boltzmann equation and we study the solution in the limit where the diffusion effects are small. This solution is employed in section \ref{forecast_sec}, where we show how to parametrize a realistic GW spectrum in the presence of the diffusion effects and we forecast the constraints posed on both the background model parameters and the diffusion and drift constants with LISA. In section \ref{conclusion_sec}, we comment on the results of our study.

\section{The stochastic GW background}
\label{SGWB_sec}
We consider a stochastic GW background as a superposition of plane waves. In the TT gauge, this can be written as
\begin{align}\label{planar_wave}
h_{ij}(t,\mathbf{x})=\int_{-\infty}^{\infty} d\nu     \int d^2\mathbf{n}\sum_A\, h^A(\nu,\mathbf{n})e^A_{ij}(\mathbf{n})e^{-2\pi i\nu(t-\mathbf{n}\cdot\mathbf{x}/c)}\,,
\end{align}
where $i$, $j$ run over spatial coordinates and the index $A$ runs over all orthogonal polarisation states, including $+$ and $\times$.
We consider only plane waves with a standard dispersion relation propagating at the speed of light. The unit vector $\mathbf{n}$ is aligned with the momentum of each wave with frequency $\nu$.

The amplitude of $h^A(\nu,\mathbf{n})$ in \eqref{planar_wave} is a stochastic variable at each frequency $\nu$ and direction $\mathbf{n}$, while the polarization tensors are defined with normalization $\sum_A e^A_{ij}(\mathbf{n})e_A^{ij}(\mathbf{n})=2$. We assume the GW background to be stationary, isotropic, Gaussian, and unpolarized, as it is predicted by the vast majority of the cosmological models:
\begin{align}\label{PSD_signal}
\langle h^A(\nu,\mathbf{n})h^{B\star}(\nu',\mathbf{n}')\rangle=\delta_{AB}\delta(\nu-\nu')\delta^{(2)}(\mathbf{n}-\mathbf{n'})S_h(\nu)\,,
\end{align}
where $S_h$ is the frequency power spectrum of the gravitational waves, while $\delta(\nu)$ and $\delta^{(2)}(\mathbf{n})$ are the 1-d and 2-d Dirac deltas, respectively.
One can relate the power spectrum in equation \eqref{PSD_signal} to the graviton distribution function in phase space $f$ via the GW energy density 
\begin{align}
\rho_{\rm GW}&=\frac{c^2\langle \dot h_{ij}\dot h^{ij}\rangle}{32\pi G}=\frac{2\pi^2 c^2}{G}\int_0^\infty d\nu\,\nu^2 S_h(\nu)\,.
\end{align}
If we define $f(E)$ to be the phase space distribution of gravitons for homogeneous distributions, with $E=2\pi\hbar \nu$ being the energy, then the energy density is found simply by integrating $Ef(E)$ over the momentum space:
\begin{align}
\label{rho_GW_in_terms_of_f}
\rho_{\rm GW}&=\frac{4\pi}{c^3}\int_0^\infty dE\,E^3 f(E)\,.
\end{align}
Comparing the last two equations, one obtains the relationship between the distribution function and power spectrum
\begin{align}
f(E)&=\frac{c^5}{32\pi^3 G\,\hbar^4} \frac{S_h(\nu)}{\nu}\,.
\end{align}
As the definition of the GW spectrum $S_h$ may vary depending on conventions, it is always convenient to introduce the logarithmic GW energy density $\Omega_{\rm GW}(\nu)$
\begin{equation}
    \frac{\rho_\text{GW}}{\rho_c}=\int_{-\infty}^\infty d\ln\nu \ \Omega_{\rm GW}(\nu)\,,
\end{equation}
where $\rho_c$ is the present critical density. We can finally relate $\Omega_{\rm GW}$ to $f$:
\begin{align}
\label{f_Omega}
f(E)=\frac{3c^5H_0^2}{512\pi^6G\hbar^4}\frac{1}{\nu^4}\Omega_\text{GW}(\nu)\,,
\end{align}
where $H_0$ is the Hubble constant today. What matters most for our later analysis is the frequency scaling in $f\sim\Omega_\text{GW}(\nu)/\nu^4$.

There is yet another way to derive the distribution $f$ in terms of $\Omega_{\text{GW}}$ which is more appealing from a theoretical perspective. The distribution function is closely related to the Wigner function, and both satisfy the Boltzmann equation. The Wigner function is the Fourier-transformed point-separated two-point function, and for gravitational waves in the  TT-gauge in flat spacetime, it is defined at spacetime point $x$ and four-momentum $p$ as 
\begin{equation}
    \mathcal{W}(x,p)=\frac{c^4}{32\pi G\hbar^2}\int\frac{d^4x'}{(2\pi\hbar)^4}\ e^{-ip.x'/\hbar}\langle h_{ij}(x+x'/2)h^{ij}(x-x'/2)\rangle\,,
\end{equation}
see \cite{fonarev1994wigner} for the extension of Wigner functions to curved spacetime. This definition in the TT-gauge has the property that the energy-momentum tensor of the gravitational waves would be given by 
\begin{equation}
\label{t_munu_in_terms_of_W}
    t_{\mu\nu}(x)=\int d^4p \ p_\mu p_\nu\mathcal{W}(x,p)\,.
\end{equation}
For another discussion of the Wigner function of the linearized gravitational field, see \cite{Garcia-Compean:2011oid}. From the equation of motion of $h_{ij}$, one can show that ignoring quantum fluctuations, $\mathcal{W}$ is non-zero only on the light cone, i.e. when $p^2=0$. Therefore, we can define the distribution function $f$ as for scalar fields \cite{groot1980relativistic}
\begin{equation}
\label{W_f}
    \mathcal{W}(x,p)=2c\ \delta(p^2)f(x,p)\,.
\end{equation}
One can see that this definition, together with \eqref{t_munu_in_terms_of_W}, gives the expression \eqref{rho_GW_in_terms_of_f} for the energy density. Now since $\mathcal{W}$ is in a sense the Fourier transform of the two-point function, one could anticipate a close relation to the power spectrum $S_h$. Indeed, using the isotropic expressions \eqref{planar_wave} and \eqref{PSD_signal} for $h_{ij}$, one can find the Wigner function in terms of the power spectrum
\begin{equation}
    \mathcal{W}(x,p)=\frac{c^5}{8\pi^2G\hbar^3}\frac{1}{p}S_h\left(\frac{cp}{2\pi\hbar}\right)\delta(p^2)\,.
\end{equation}
This, together with the definition of the distribution function \eqref{W_f}, gives the relation between $f$ and $S_h$ that we found earlier by just comparing the energy density. However, knowing that $\mathcal{W}$ is a Wigner function ensures us that the distribution $f$ satisfies the Boltzmann equation. Therefore, any effective modification to the Boltzmann equation from quantum gravity would show up in the transport equation for $f$, which is related to $\Omega_{\text{GW}}$ via \eqref{f_Omega}.

\section{Modified transport equation for the distribution function}
\label{Boltzmann_sec}

In the absence of interactions and collisions, the phase space distribution of the gravitational waves, $f$, satisfies a Boltzmann equation
\begin{equation}
     L[f]\equiv p^\mu\frac{\partial}{\partial x^\mu}f+\Gamma^\sigma_{\mu\nu}p_\sigma p^\mu\frac{\partial}{\partial p_\nu}f=0\,.
\end{equation}
The term proportional to the Christoffel symbol is what gives the Hubble friction in FRW spacetime. 

As mentioned in the introduction, many proposals motivated by quantum gravity modify the dispersion relation $p^2=0$ and hence break Lorentz invariance. However, motivated by stringent constraints on Lorentz violation \cite{Vasileiou2013vra,Ellis2018lca,LHAASO2024lub,Piran2023xfg,MAGIC2007etg,HESS2011aa}, we maintain the light-cone propagation of gravitons and preserve $p^2=0$. Then, at an effective level, any quantum gravity effect must appear in the geodesic equation for $p$ instead, and the modification should be such that $p^2=0$ is preserved by the new dynamics. In general, a modification to the geodesic equation includes a drift and a noise term, where the drift has to be a vector and the covariance matrix of the noise has to be a tensor. Assuming Lorentz invariance and no coupling to curvature, in \cite{albertini2025stochastic} it was shown that the correction can be described by two free parameters $\kappa_{1,2}$:
\begin{align}
\label{affine_normalization}
    dx^\mu &= p^\mu d\lambda \,,\\\label{massless_SDE_ansatz}
    dp^\mu +\Gamma ^\mu_{\alpha \beta} p^{\alpha} p^\beta d\lambda &= \kappa_2 p^\mu d\lambda + p^\mu dW_\lambda \ ,\\
    \big\langle dW_\lambda\big\rangle = 0 \ , &\quad \big\langle dW_\lambda^2\big\rangle = 2\kappa_1d\lambda \ ,
\end{align}
where $dW_\lambda$ is a Gaussian noise with its strength controlled by $\kappa_1$. More geometrically, a single graviton would experience a random walk in its one-dimensional direction of motion on the light cone on top of a drift toward higher or lower frequencies controlled by $\kappa_2$. Note that $\kappa_1$ controls the variance of the noise and must be non-negative, while $\kappa_2$ can be positive or negative depending on whether the microscopic effect causes a blue or red tilt.

The solution to the stochastic geodesic equation \eqref{massless_SDE_ansatz} in an FRW spacetime with scale factor $a$ is given by
\begin{equation}
\label{solution_to_stoch_geodesic}
    E=\frac{E_0}{a} e^{(\kappa_2-\kappa_1)\Delta\lambda+W_{\Delta\lambda}}\,.
\end{equation}
Note that $E$ is the physical energy at scale factor $a$, i.e. $p^0=E/a$, and $E_0$ is the energy at present time. $\Delta\lambda$ is the total elapsed affine time. As already pointed out in \cite{philpott2009energy}, the normalization of the affine time is not arbitrary but set by \eqref{affine_normalization} and the knowledge of physical momentum. $W_{\Delta\lambda}$ is the integral of the noise and is a random walk with variance $2\kappa_1\Delta\lambda$.

The above solution recovers the $1/a$ redshifting of the particles for zero diffusion while clearly showing that the strength of the diffusion and drift is controlled by the dimensionless quantity $\kappa_{1,2}\Delta\lambda$. Note that given a frame, high-energy particles traverse the same distance in a shorter affine time. So low-frequency particles would experience stronger diffusion and drift. At first, this appears counter-intuitive, as we are used to the idea of larger frequencies probing smaller scales. However, frequency or the number of oscillations of a propagating wave is frame-dependent, while the elapsed affine time defined above is Lorentz invariant. The elapsed affine time measures the spacetime area swept by one de-Broglie wavelength of the particle \cite{philpott2009energy}. The longer the wavelength, the larger this area. So a quantum-gravity effect that acts as independent stochastic kicks at Planck-scale volumes would add up with a standard deviation proportional to the square root of the spanned spacetime area of a wave. 

At the level of distribution of particles, this corresponds to the following modified Boltzmann equation
\begin{equation}
    p^\mu\frac{\partial}{\partial x^\mu}f+\Gamma^\sigma_{\mu\nu}p_\sigma p^\mu\frac{\partial}{\partial p_\nu}f=\big(3\kappa_1-\kappa_2\big)\frac{1}{E}\frac{\partial}{\partial E}\big(E^2f\big)+\kappa_1\frac{1}{E}\frac{\partial}{\partial E}\Big(E^3\frac{\partial}{\partial E}f\Big),
\end{equation}
reflecting the fact that a peaked distribution of gravitons broadens and drifts over time. For example, at the background level in an FRW metric with conformal time $\eta$ and scale factor $a$, we have $p^0=E/a$. So the equation becomes
\begin{equation}\label{eq:eqdiff_etaq}
    \frac{1}{a}\frac{\partial}{\partial \eta}f-\frac{1}{a}E\mathcal{H}\frac{\partial}{\partial E}f=\big(3\kappa_1-\kappa_2\big)\frac{1}{E^2}\frac{\partial}{\partial E}\big(E^2f\big)+\kappa_1\frac{1}{E^2}\frac{\partial}{\partial E}\Big(E^3\frac{\partial}{\partial E}f\Big)\,,
\end{equation}
where $\mathcal{H}=a'/a$. Note that in order to match with the notation used in \cite{philpott2009energy}, we have to identify $k_1=\kappa_1$, $k_2=\kappa_2-3\kappa_1$, and $\rho_t\propto a^3E^2f$.

In the next section, we show the analytic solutions to this equation for a general expansion history.

\section{Analytic Solution}
\label{analytic_sol_sec}

Here we present two heat kernel solutions to the covariant diffusion equation and study their behavior. Introducing the comoving energy as $\epsilon=aE$ and a new time variable $T$ as $dT=a^2d\eta$, the left side of equation \eqref{eq:eqdiff_etaq} simplifies to $a\partial_Tf$. Following the steps detailed in Appendix \ref{app:analytic_sol}, one can construct two heat kernels for this equation
\begin{align}
    \boxed{\label{h_closed_form}
    K_\pm(T,\epsilon;\epsilon_i)=\frac{1}{\epsilon_i}\exp\left(-\frac{\epsilon+\epsilon_i}{\kappa_1(T-T_i)}\right)\Big(\frac{\epsilon_i}{\epsilon}\Big)^{3-\frac{\kappa_2}{2\kappa_1}}\frac{\sqrt{\epsilon\epsilon_i}}{\kappa_1(T-T_i)}I_{\pm(1-\frac{\kappa_2}{\kappa_1})}\Big(\frac{2\sqrt{\epsilon\epsilon_i}}{\kappa_1(T-T_i)}\Big)}\,,
\end{align}
where $I_\alpha$ is the modified Bessel function. The heat kernels satisfy the initial condition 
\begin{align}
    K_\pm(T_i,\epsilon;\epsilon_i)=\delta(\epsilon-\epsilon_i)\,.
\end{align}
One can check this using the zeroth-order term in the asymptotic expansion of the Bessel function. When $\kappa_1\Delta T$ is small (compared to energy scales $\epsilon$ and $\epsilon_i$), $K_\pm$ is indeed a function of $\epsilon$ peaked at $\epsilon_i$ and integrates to 1. So in the limit $\kappa_1\Delta T\rightarrow0$, it becomes the delta function, as expected.

The above heat kernels provide two solutions to equation \eqref{eq:eqdiff_etaq} given an initial condition at some initial time $T_i$ is provided by $f(T_i,\epsilon_i)$ for all $\epsilon_i>0$. These can be written as the following integral
\begin{equation}
\label{integral_solution}
\boxed{f(T,\epsilon)=\int_{0}^{\infty}d\epsilon_i\ f(T_i,\epsilon_i)K_\pm(T,\epsilon;\epsilon_i)}\,.
\end{equation}
This is similar to a convolution but with the difference that the kernel is not a function of $\epsilon-\epsilon_i$ because the diffusion operator is not translationally invariant in $\epsilon$.

However, the fact that we have found two different solutions to our diffusion equation is puzzling at the start. Mathematically, this happens because $\epsilon=0$ is a singular point in the original PDE \eqref{eq:eqdiff_etaq}, and hence there is no uniqueness of solutions. Nevertheless, there must be only one physically acceptable solution. We can show that this is indeed the case by demanding the total particle number to be finite and conserved.

From the solution \eqref{solution_to_stoch_geodesic} to the stochastic geodesic equation, we know that the underlying particle model conserves the particle number; i.e. no particle gets lost at the tip of the light cone or zero frequency. So we expect the distribution of the particles to be conservative as well. In terms of the heat kernels, we demand $\int_0^\infty d\epsilon\ \epsilon^2K_\pm$ to be equal to $\epsilon_i^2$ at all times.

In Appendix \ref{app:analytic_sol}, we show that for the given heat kernels, we have
\begin{equation}
\label{particle_number_for_heat_kernels}
\int_0^\infty d\epsilon\ \epsilon^2K_+=\left\{
\begin{aligned}
&\epsilon_i^2\,,\hspace{43mm} \text{if }\kappa_2/\kappa_1\in\{1,2,3,...\} \\&
\epsilon_i^2\Big(1-\frac{\Gamma(1-\kappa_2/\kappa_1,\frac{\epsilon_i}{\kappa_1\Delta T})}{\Gamma(1-\kappa_2/\kappa_1)}\Big)\,,\ \ \text{if }\kappa_2/\kappa_1\notin\{1,2,3,...\}
\end{aligned}
\right.\,,
\end{equation}

\begin{equation}
\int_0^\infty d\epsilon\ \epsilon^2K_-=\left\{
\begin{aligned}
&\epsilon_i^2\,,\hspace{43mm} \text{if }\kappa_2>0 \\&
\epsilon_i^2\Big(1-\frac{\Gamma(1-\kappa_2/\kappa_1,\frac{\epsilon_i}{\kappa_1\Delta T})}{\Gamma(1-\kappa_2/\kappa_1)}\Big)\,,\ \ \text{if }\kappa_2/\kappa_1\in\{0,-1,-2,...\}
\\& \infty \,,\hspace{43mm} \text{else}
\end{aligned}
\right.\,.
\end{equation}
So for $\kappa_2>0$, $K_-$ gives a physically acceptable heat kernel. For natural number values of $\kappa_2/\kappa_1$, $K_+$ is also acceptable, but then it actually coincides with $K_-$ because for integer values of $\alpha$ we have $I_{-\alpha}(z)=I_\alpha(z)$. So for $\kappa_2>0$, we have found a unique physically acceptable solution.

For $\kappa_2\leq0$, however, none of the above heat kernels are physically allowed, as they either gradually lose particles at $\epsilon=0$ or give a divergent particle number. Note that this does not mean that the massless diffusion model is only valid for $\kappa_2>0$, but that we have not been able to find the correct heat kernel with our approach for solving the diffusion equation.

\begin{figure}[h!]
    \centering
    \begin{subfigure}[b]{0.495\textwidth}
        \centering
        \includegraphics[width=\textwidth]{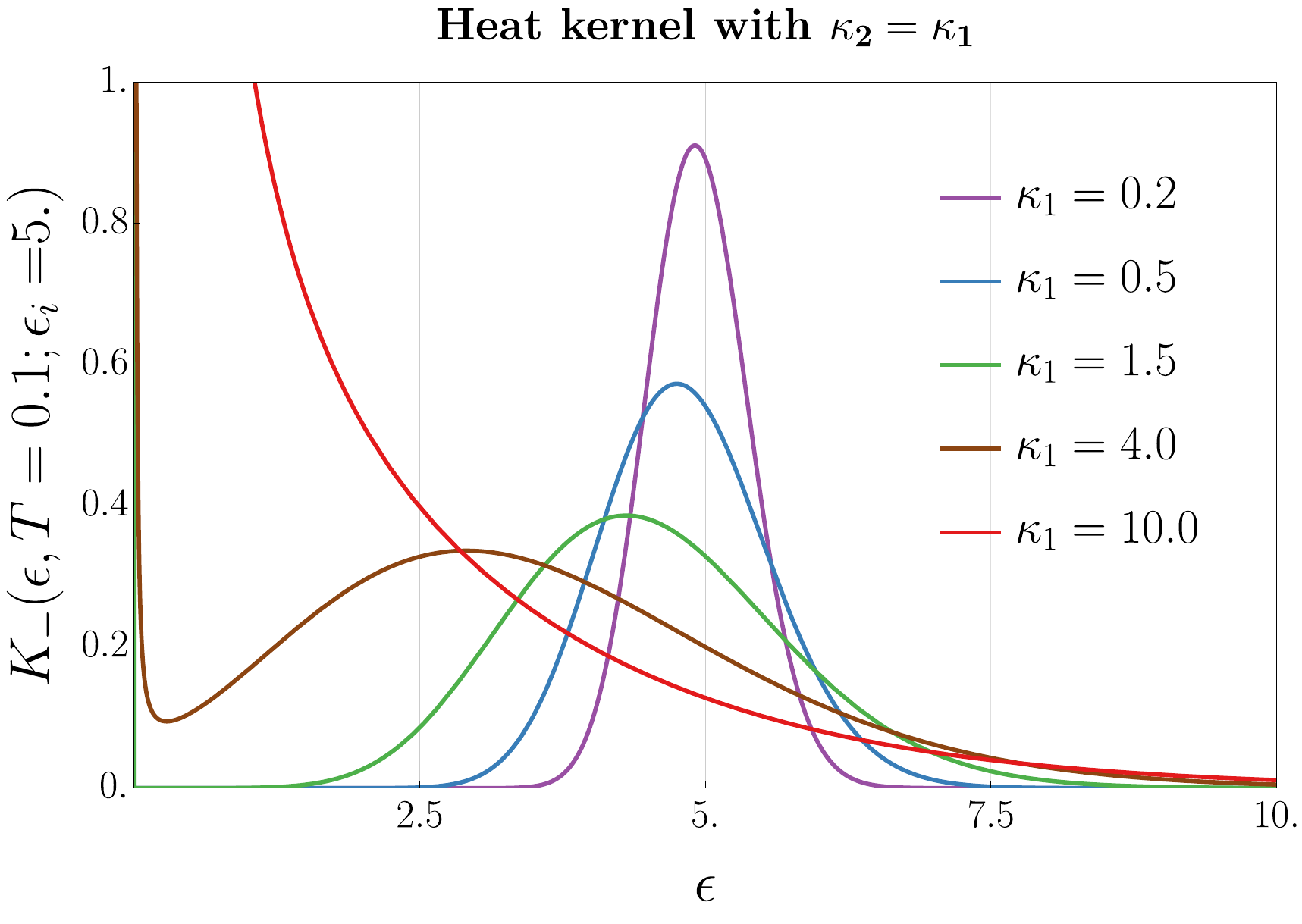}
        \caption{}       \label{k2=k1}
    \end{subfigure}
    \hfill
    \begin{subfigure}[b]{0.495\textwidth}
        \centering
        \includegraphics[width=\textwidth]{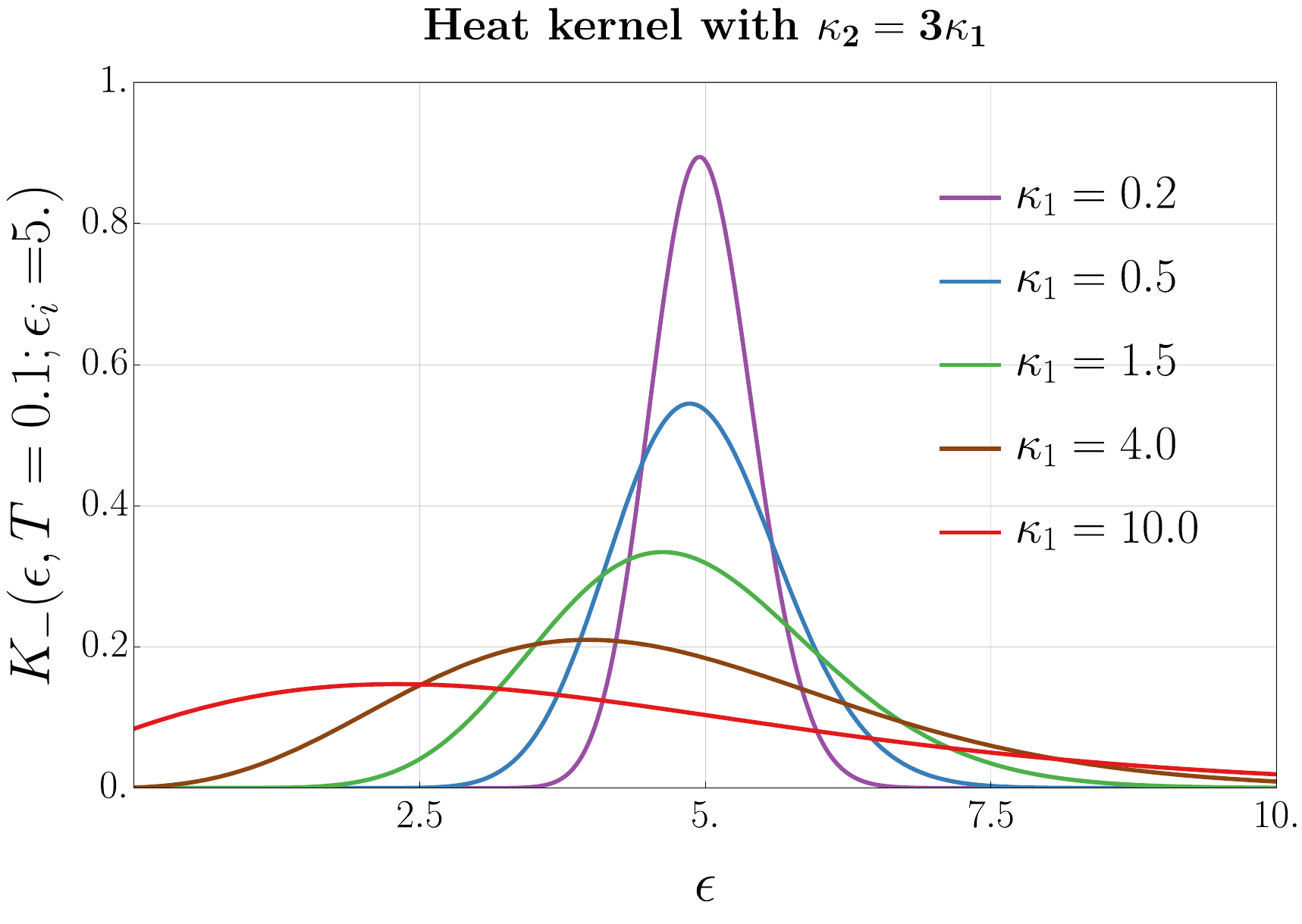}
        \caption{}\label{w_plot}
    \end{subfigure}
    \caption{\fontsize{9}{11}\selectfont The physical heat kernel $K_-$ for a time laps of $\Delta T=0.1$, and different values of $\kappa_{1,2}$ in arbitrary units. $\epsilon_i=5$, so all the kernels are evolved from an initial delta function at $\epsilon=5$. A larger value for $\kappa_2$ causes a relative drift toward higher energies.}
    \label{h_plots}
\end{figure}

Nevertheless, it is important to note that the heat kernels $K_\pm$ are still valid for $\kappa_2\leq0$ if we avoid small frequencies; i.e. small values of $\frac{\epsilon}{\kappa_1\Delta T}$. To back up this claim, notice that the difference between $I_{-\alpha}(z)$ and $I_\alpha(z)$ is proportional to the modified Bessel function of the second kind. Therefore, the difference $K_+-K_-$ is exponentially small for large values of $\frac{\epsilon}{\kappa_1\Delta T}$, and becomes noticeable only as we approach $\frac{\epsilon}{\kappa_1\Delta T}\rightarrow0$, where the power law behaviors of $K_+$ and $K_-$ are different. In particular, for $\kappa_2>0$, $K_-$ behaves like $\epsilon^{\kappa_2/\kappa_1-3}$ at small frequencies. This can be seen in figure~\ref{h_plots} for two different rations of $\kappa_2/\kappa_1$. Also note that the curves for increasing values of $\kappa_1$ could be equivalently seen as increasing time $\Delta T$. For explicit expressions for the power-law behavior of the two kernels close to zero frequency see Appendix \ref{app:analytic_sol}.

\subsection{Perturbative solution for small $\kappa_{1,2}$}

Although we have the full analytic solution of the massless diffusion equation given by \eqref{integral_solution}, it would be computationally expensive to perform the integral for every value of $\kappa_1,\ \kappa_2$, and $\epsilon$ in a numerical forecast. Moreover, all that matters for the sake of the Fisher forecast (which will be treated in detail in the next section) is the Taylor expansion of the solution for $f$ up to the first order in $\kappa_1$ and $\kappa_2$.

Therefore, we expand our analytic solution to first order in $\kappa_{1,2}$. More precisely, the small dimensionless parameter over which we perform the Taylor expansion is $\frac{\kappa_1\Delta T}{\epsilon}$. So this can be interpreted as an approximation on small enough time scales or at large enough frequencies as well. For primordial gravitational waves, $\Delta T\approx a_0^2\eta_0/5$, where $\eta_0$ is the conformal age of the universe. This time-of-flight is dominated by the late-universe matter-dominated evolution. For a mHz experiment like LISA, $\frac{\kappa_1\Delta T}{\epsilon}$ is of order 1 for $\kappa_1\sim10^{-55}\text{kg}\,\text{m}^2\,\text{s}^{-3}$. So for our forecast, we restrict attention to values of $\kappa_{1,2}$ smaller than this. Otherwise, for values of $\kappa_{1,2}$ comparable to this number or larger, the effect on the spectrum would be very drastic, as can be seen with the red curves in figure \ref{h_plots}. Note that the regime of small $\frac{\kappa_1\Delta T}{\epsilon}$ is also the regime in which $K_{\pm}$ coincide, and so the choice of the heat kernel becomes irrelevant. Also, the aforementioned issue of invalidity of the heat kernels for $\kappa_2\leq0$ is irrelevant for the same reason.

In the regime where $\frac{\epsilon}{\kappa_1\Delta T}\gg1$, we need the asymptotic expansion of the modified Bessel function
\begin{equation}
    I_\alpha(z) = \frac{e^z}{\sqrt{2 \pi z}} \Big(1 - \frac{4 \alpha^2 - 1}{8 z}+\mathcal{O}(1/z^2)\Big)\,.
\end{equation}
Since $\alpha=\pm(1-\frac{\kappa_2}{\kappa_1})$ only appears as $\alpha^2$, both $K_{\pm}$ have the same asymptotic expansion as expected. We then plug the above expansion in $K_{\pm}$, and change the heat kernel integration variable in \eqref{integral_solution} from $\epsilon_i$ to $y$ using $\sqrt{\epsilon_i}=\sqrt{\epsilon}+y\sqrt{\kappa_1\Delta T}$. After Taylor-expanding the integrand around $\kappa_1\Delta T=0$, the integrand becomes of the form $\exp(-y^2)$ times a series in $y$. So in the perturbative regime of small diffusion, the heat kernel integral \eqref{integral_solution} becomes a simple Gaussian integral, giving
\begin{equation}\label{eq:f_perturbed_final}
f(T,\epsilon)=f_i(\epsilon)+\left[2\left(3-\frac{\kappa_2}{\kappa_1}\right)\frac{1}{\epsilon}f_i(\epsilon)+\left(6-\frac{\kappa_2}{\kappa_1}\right)\partial_\epsilon f_i(\epsilon)+\epsilon\partial^2_\epsilon f_i(\epsilon)\right]\kappa_1\Delta T+\mathcal{O}\left(\left(\frac{\kappa_1\Delta T}{\epsilon}\right)^2\right),
\end{equation}
where $f_i(\epsilon):=f(T_i,\epsilon)$. Given that we know the initial power spectrum analytically, this is very simple to evaluate numerically.

\section{Forecast of $\kappa_{1,2}$ estimation with LISA}
\label{forecast_sec}

We aim to forecast the observability of a stochastic GW background by estimating the posterior distribution over the model parameters $\theta$ that characterize the power spectrum $\Omega_{\rm GW}(\nu;\theta)$. Specifically, we define $\theta=\{\theta_0,\kappa_1,\kappa_2\}$, where $\theta_0$ is the set of parameters that model the unperturbed spectrum, i.e. in the absence of the diffusion effect, $\Omega_{0}(\nu;\theta_0)$, and $\kappa_{1,2}$ are the newly introduced diffusion and drift parameters that appear explicitly in equation \eqref{eq:f_perturbed_final}.
For forecasting purposes, we specify the fiducial value for the model parameters $\bar\theta=\{\bar\theta_0,\bar\kappa_1,\bar\kappa_2\}$. Provided that $\bar\kappa_1,\bar\kappa_2 \ll 1$ in our units, we express equation~\eqref{eq:f_perturbed_final} as
\begin{align}
f(\epsilon;\theta)&\simeq f_0(\epsilon;\theta_0)+\kappa_1 f_1(\epsilon;\theta_0)+\kappa_2 f_2(\epsilon;\theta_0)\,,\nonumber\\
f_{1,2}(\epsilon;\theta_0)&=\frac{\partial f}{\partial \kappa_{1,2}}_{\Big |_{\kappa_{1,2}=\bar\kappa_{1,2}}}\,.
\end{align}
At the present time where $a_0=1$, the frequency is given by $\nu=\frac{\epsilon}{2\pi\hbar}$. Using equation \eqref{f_Omega}, the above equation translates into an expression for the GW energy density:
\begin{align}\label{eq:Omega_GW_Taylored}
\Omega_{\rm GW}(\nu;\theta)&\simeq \Omega_{0}(\nu;\theta_0)+\kappa_1 \Omega_{1}(\nu;\theta_0)+\kappa_2 \Omega_{2}(\nu;\theta_0)\,,\nonumber\\
\Omega_{i}(\nu;\theta_0)&=\frac{\partial \Omega_{\rm GW}(\nu;\theta)}{\partial \kappa_{i}}_{\Big |_{\kappa_{i}=\bar\kappa_{i}}}\,.
\end{align}
A stochastic GW background search with LISA exploits the fact that the instrument provides three data streams with (approximately) uncorrelated instrumental noise and a correlated signal in the presence of a GW background. It can be shown that the three correlators obtained between the LISA data streams follow a Gaussian distribution.
Since the correlators between the data streams are proportional to the GW power spectrum and therefore the GW energy density, the likelihood function in the LISA correlators can be written as a function of the theoretical parameters $\theta$. We report the explicit form of the likelihood function without reproducing the derivation, which can be found in \cite{LISACosmologyWorkingGroup:2022kbp}: 
\begin{align}\label{eq:Likelihood}
\mathcal{L}(\theta)&=\frac{1}{\mathcal{N}}e^{-\frac{1}{2}\chi^2(\theta)}\,,\nonumber\\
\chi^2(\theta)&=T_{\rm obs}\int_{\nu_{\rm min}}^{\nu_{\rm max}} d\nu\,\frac{4\pi\,(\Omega_{\rm GW}(\nu;\theta)-\Omega_{\text{measured}}(\nu))^2}{\Omega_{n}^2(\nu)}\,,
\end{align}
where $\mathcal{N}$ is a normalization constant and $\Omega_{n}(\nu)$ is the sensitivity curve for an isotropic stochastic GW background, obtained from the combined LISA data channels and for a total time of observation $T_{\rm obs}$. The frequency bandwidth for LISA is assumed to be between $\nu_{\rm min}=0.1$mHz to $\nu_{\rm max}=100$mHz.
The computation of $\Omega_{n}(\nu)$ follows the same steps of the derivation of the noise curve in \cite{LISACosmologyWorkingGroup:2022kbp} for an isotropic GW background that leads to their equation (4.43).
The posterior distribution on the parameters $\theta$ can be obtained via the Bayes theorem given a prior function $\Pi(\theta)$
\begin{align}
P(\theta)=\Pi(\theta)\mathcal{L}(\theta)\,.
\end{align}
The explicit evaluation of the likelihood function in \eqref{eq:Likelihood} is in general computationally expensive, given the high dimensionality of the space of parameters and the non-trivial dependence of the signal on these parameters in the LISA data. For forecasting purposes, approximate techniques, such as the Fisher method outlined in appendix \ref{app:Fisher}, provide the same quantitative and qualitative information as simulation-based approaches. They allow us to reliably forecast the degree of detectability and estimate the theoretical parameters.
In appendix \ref{app:Fisher}, we analytically derive the covariance matrix for the theoretical parameters in the two specific models considered in the following subsection. This provides insight into how diffusion effects can correlate with the parameters of the unperturbed GW background model.

\subsection{Diffusion on first order phase transitions and PBH spectra}\label{subsec:PT_PBH}

\begin{figure}[h!]
    \centering
    \begin{subfigure}[b]{0.49\textwidth}
        \centering
        \includegraphics[width=\textwidth]{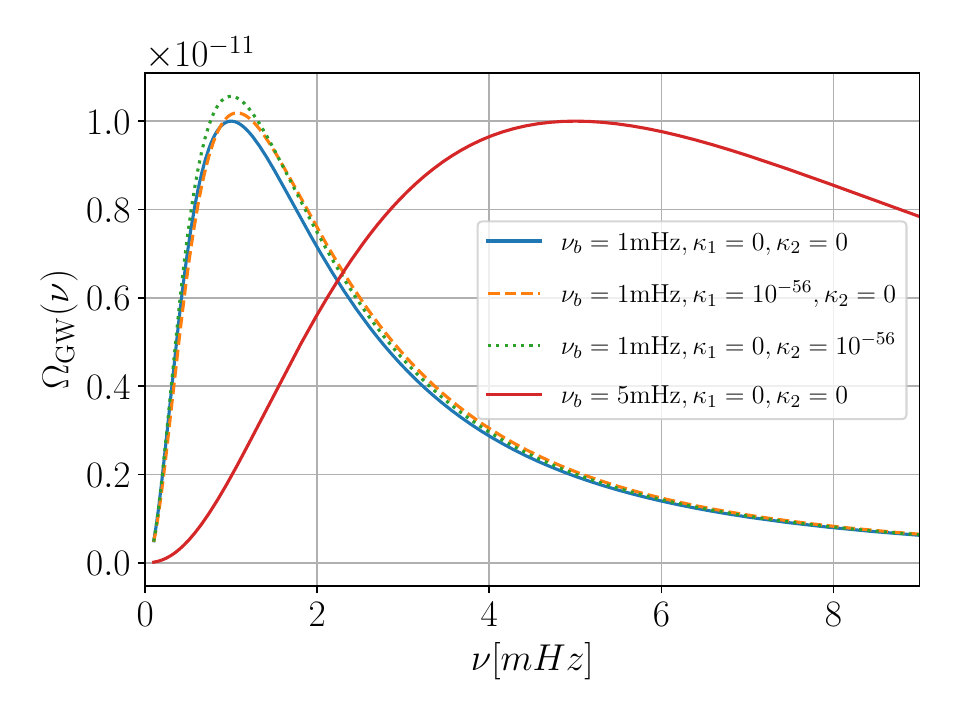}
    \end{subfigure}
    \hfill
    \begin{subfigure}[b]{0.5\textwidth}
        \centering
        \includegraphics[width=\textwidth]{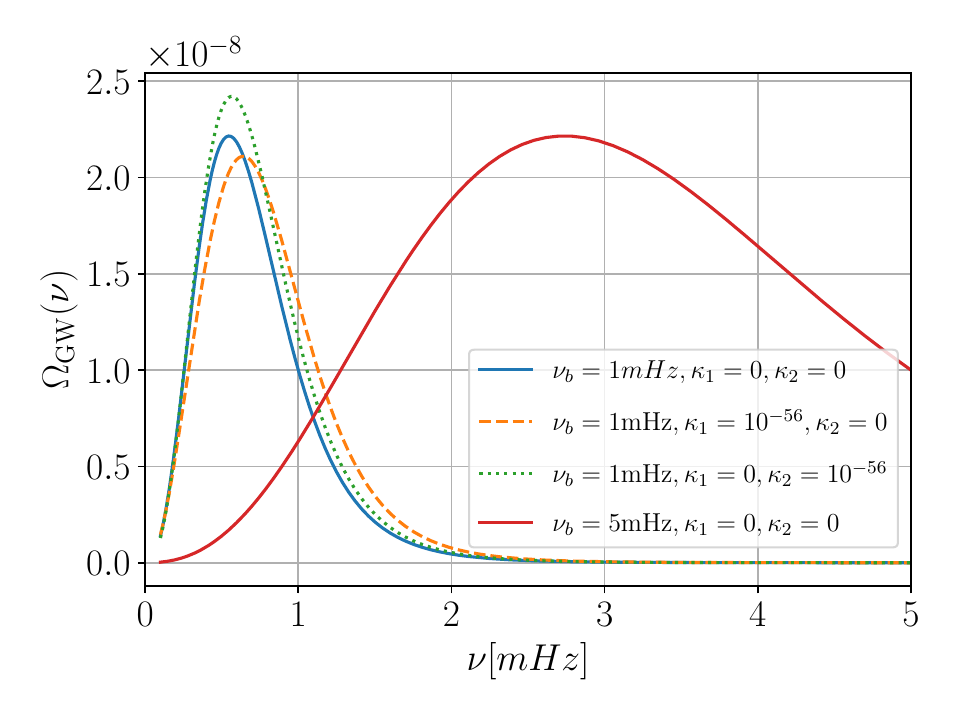}
    \end{subfigure}
    \caption{\fontsize{9}{11}\selectfont Plot of the GW energy density $\Omega_{\rm GW}(\nu)$ of a stochastic background from a first-order phase transition (left) and from primordial black holes (right). 
    We show in dashed lines how the diffusion effect broadens and shifts the unperturbed spectra for different values of $\kappa_{1,2}$ in $\text{kg}\,\text{m}^2\,\text{s}^{-3}$.
    We parametrize the unperturbed spectra as broken power laws as in equation \eqref{eq:OmegaGW_PT}, with different fiducial values for the parameters.} 
    \label{fig:Omega_PT_PBH}
\end{figure}

We apply the outlined formalism to the specific case of the GW background from a strong first-order phase transition and the background produced in a PBH scenario.
In particular, we adopt the broken power-law parametrization employed in \cite{Caprini:2024hue} as it effectively describes the spectral shape of GW backgrounds generated by first-order phase transitions:
\begin{align}\label{eq:OmegaGW_PT}
\Omega_{0}\left(\nu\right)=\Omega_b\left(\frac{\nu}{\nu_b}\right)^{n_1}\left[\frac{1}{2}+\frac{1}{2}\left(\frac{\nu}{\nu_b}\right)^{a_1}\right]^{\frac{n_2-n_1}{a_1}}\equiv\Omega_b\,\psi_0(\nu;\nu_b)\,,
\end{align}
where we set $n_1=2.4$, $n_2=-2.4$, and $a_1=1.2$ as it has been found from the numerical fit to a strong phase transition \cite{Caprini:2024hue,Lewicki:2022pdb}, while we leave the overall energy density amplitude $\Omega_b$ and the characteristic frequency $\nu_b$ as free parameters. In our notation, this means that the theoretical parameters to be inferred are $\theta=\{\Omega_b,\nu_b,\kappa_1,\kappa_2\}$.
We employ the same parametrization to model the PBH background, which serves as our second toy model, with different values for $n_1=3,\ n_2=-8$, and $a_1=2$.
The choice of $n_1$ is motivated by the fact that for a physical GW energy density, a $\sim \nu^3$ scaling is expected at low frequencies \cite{Espinosa:2018eve}. We note that the first-order phase transition energy density has a different low-frequency scaling since the numerical fit in \cite{Lewicki:2022pdb} is performed on a simulation that does not take the Hubble expansion into account. The parameters $n_2$ and $a_1$ are chosen to mimic the high-frequency behavior of the GW energy density displayed in figure 2 of \cite{Bartolo:2018rku}.
The perturbed $\Omega_{\rm GW}$ in \eqref{eq:Omega_GW_Taylored} can be rewritten as
\begin{align}
\Omega_{\rm GW}(\nu;\theta)&\simeq \Omega_b\left[\psi_0(\nu;\nu_b)+\kappa_1 \psi_1(\nu;\nu_b)+\kappa_2 \psi_2(\nu;\nu_b)\right]\,.
\end{align}
Note that from equations \eqref{eq:Omega_GW_Taylored} and \eqref{eq:OmegaGW_PT}, the functions $\psi_{1,2}(\nu;\nu_b)=\Omega_{1,2}(\nu;\nu_b)/\Omega_b$ do not depend on $\Omega_b$.
The plot of $\Omega_{\rm GW}(\nu;\nu_b)$ is shown in figure \ref{fig:Omega_PT_PBH} for different choices of $\Omega_b,\, \nu_b,\, \kappa_1,$ and $\kappa_2$ for both models under consideration. Qualitatively, the presence of a diffusion term with $\kappa_1>0$ results in a blue-shift and broadening of the GW energy density, while $\kappa_2>0$ sharpens any peak present in the unperturbed spectrum.

\begin{figure}[htbp]
  \centering
  \begin{subfigure}[t]{0.495\textwidth}
    \includegraphics[width=\linewidth]{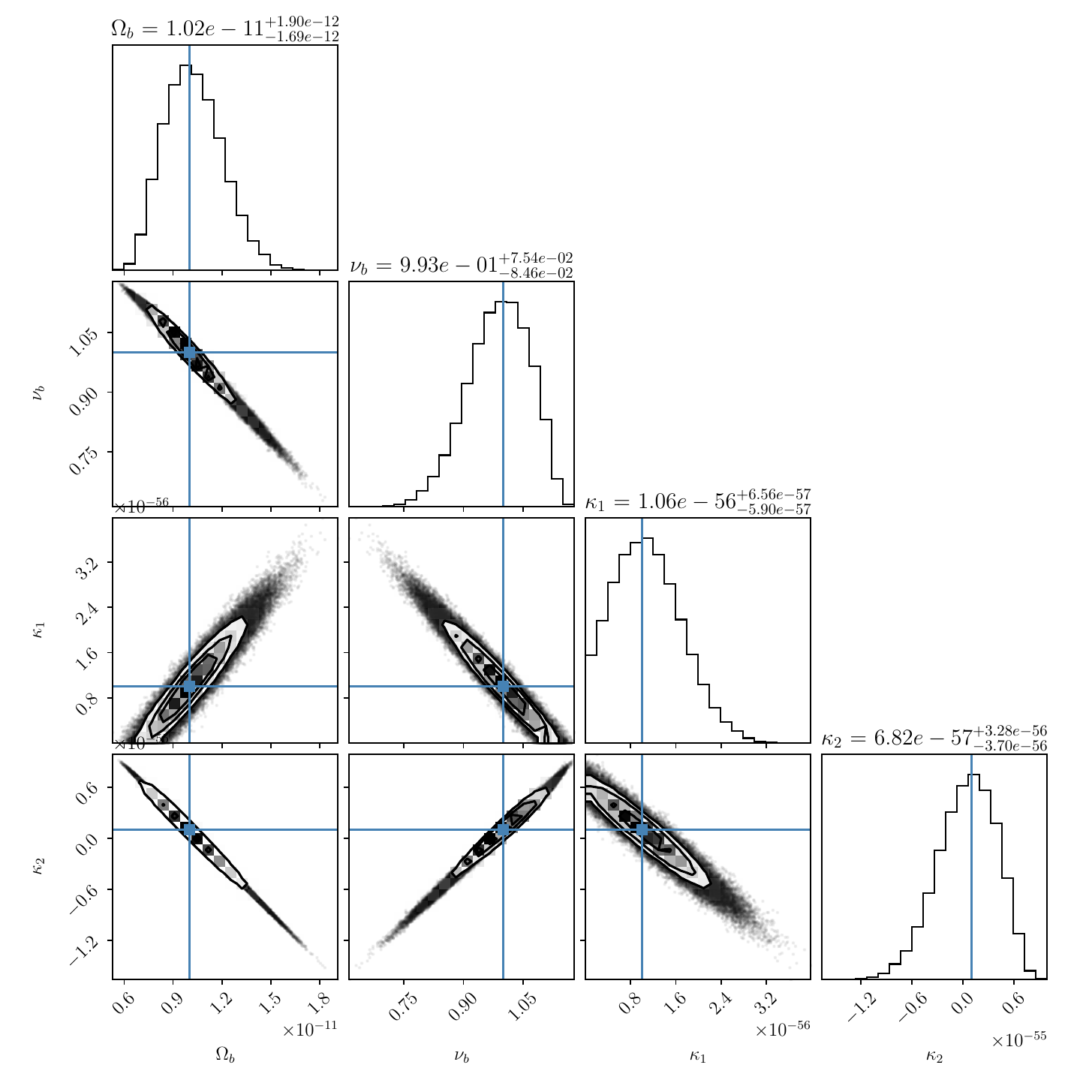}
  \end{subfigure}
  \hfill
  \begin{subfigure}[t]{0.495\textwidth}
    \includegraphics[width=\linewidth]{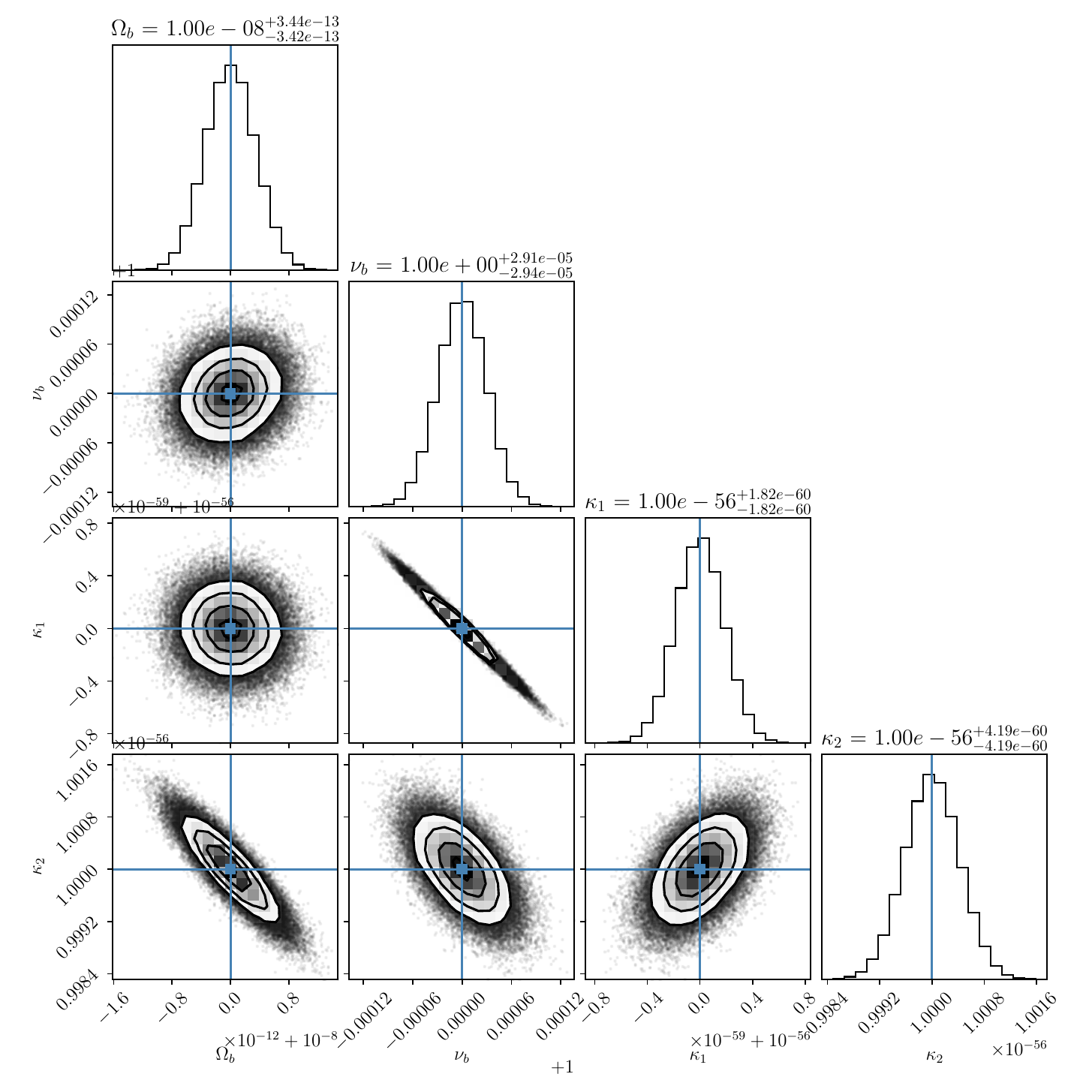}
  \end{subfigure}

  \vspace{0.5em}

  \begin{subfigure}[t]{0.495\textwidth}
    \includegraphics[width=\linewidth]{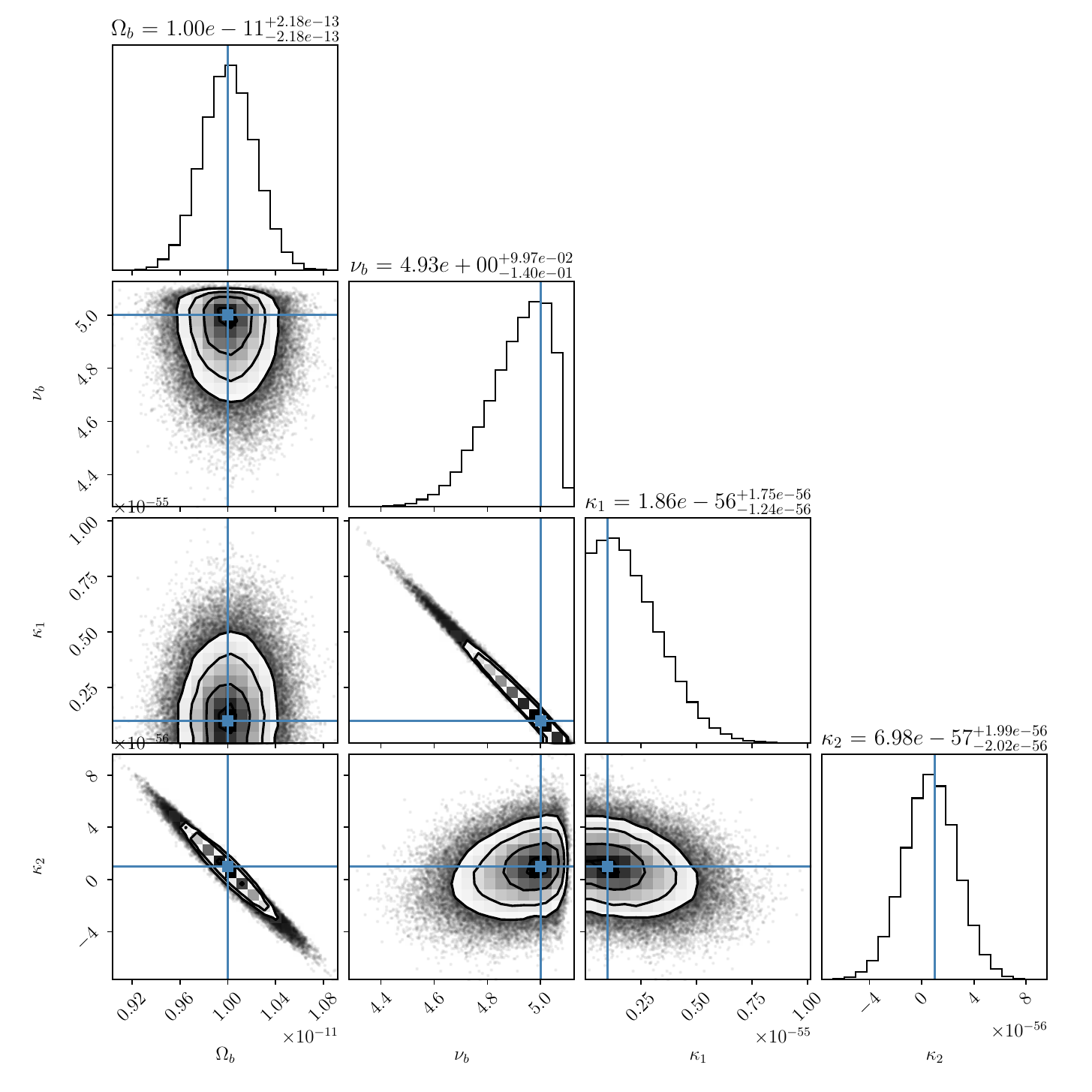}
  \end{subfigure}
  \hfill
  \begin{subfigure}[t]{0.495\textwidth}
    \includegraphics[width=\linewidth]{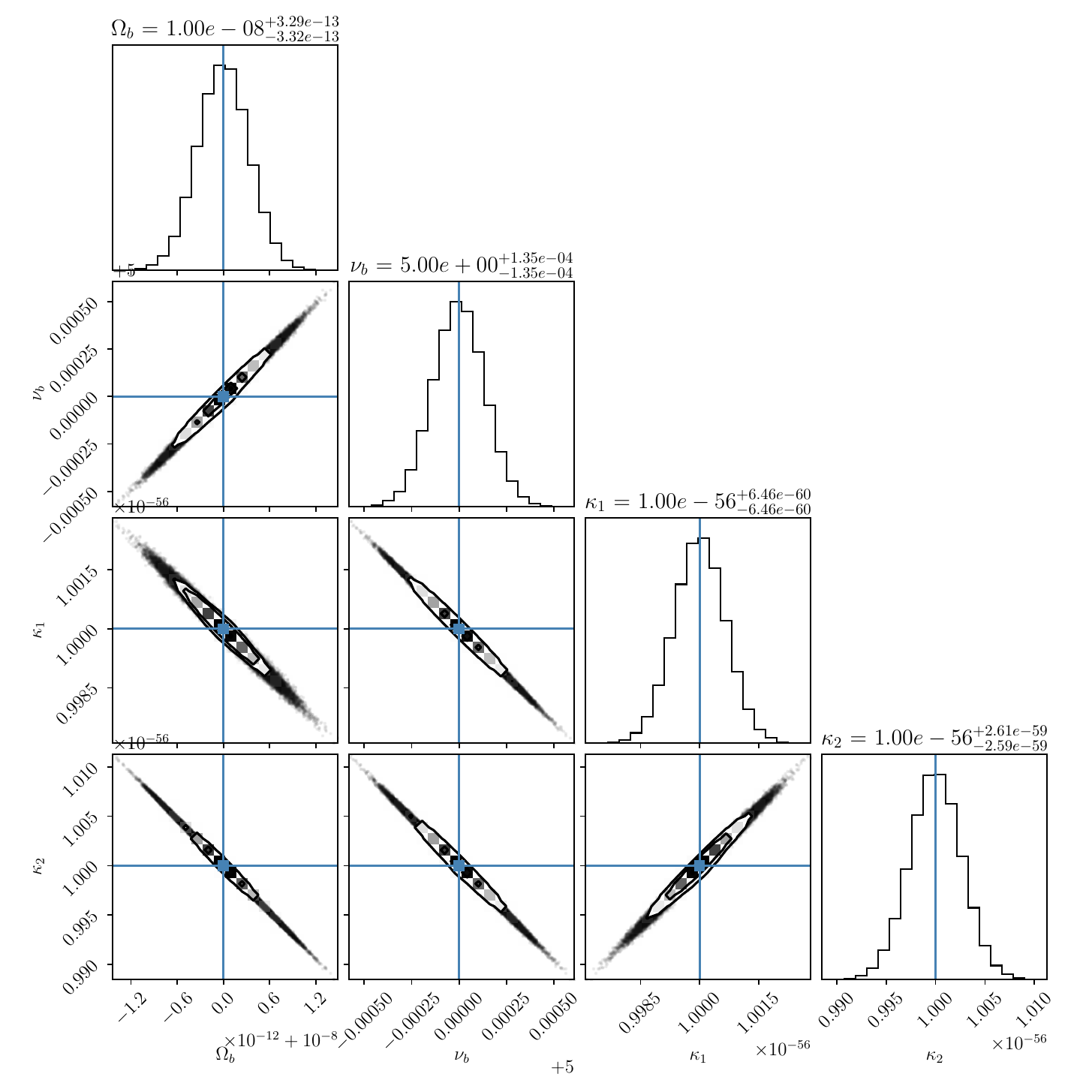}
  \end{subfigure}

  \caption{Corner plot of the forecast full posterior distribution $P(\Omega_b,\nu_b,\kappa_1,\kappa_2)$ assuming a flat prior and an effective total time of observation with LISA of $T_{\rm obs}=5$ years. Top left: a first-order phase transition scenario with fiducial values $\bar\Omega_b=10^{-11}$ and $\bar\nu_b=$1mHz. Top right: a PBH scenario with fiducial values $\bar\Omega_b=10^{-8}$ and $\bar\nu_b=$1mHz. Bottom left: a first-order phase transition scenario with fiducial values $\bar\Omega_b=10^{-11}$ and $\bar\nu_b=$5mHz. Bottom right: a PBH scenario with fiducial values $\bar\Omega_b=10^{-8}$ and $\bar\nu_b=$5mHz. Fiducial values of $\bar\kappa_1=10^{-56}\text{kg}\,\text{m}^2\,\text{s}^{-3}$ and $\bar\kappa_2=10^{-56}\text{kg}\,\text{m}^2\,\text{s}^{-3}$ are assumed in all the four cases.}
  \label{fig:corner_plot}
\end{figure}

Figure \ref{fig:corner_plot} shows the full corner plot for the forecast posterior distribution $P(\Omega_b,\nu_b,\kappa_1,\kappa_2)$ for the theoretical parameters in the first-order phase transition and the PBH models. We assume flat priors on the parameters and impose that $\Omega_b,\, \nu_b,$ and $\kappa_1$ are positive.
We find that if the spectrum has a strong enough signal in the LISA frequency band, all the parameters can be constrained, and there are no fundamental degeneracies between them.
The effect of the diffusion parameter $\kappa_1$ is anti-correlated with the characteristic frequency $\nu_b$, as they both change the peak frequency. Similarly, $\kappa_2$ always shows some degree of anti-correlation with the overall GW energy density amplitude $\Omega_b$, since they both enhance the height of the peak as they increase.
The posterior distribution for the first-order phase transition case with $\bar\nu_b=1$mHz (top-left panel in figure~\ref{fig:corner_plot}) and the posterior distribution for the PBH scenario with $\bar\nu_b=5$mHz (bottom-right panel) show stronger correlations between the parameters compared to the same models with $\bar\nu_b=5$mHz and $\bar\nu_b=1$mHz respectively. The reason for these correlations has to be attributed to the fact that the Fisher matrix is built by nontrivial combinations of $\psi_1(\nu;\nu_b),\,\psi_2(\nu;\nu_b)$, and $\frac{\partial \psi_0(\nu;\nu_b)}{\partial\nu_b}$, along with the instrumental noise curve.
We emphasize that even if the GW background may not have a peak that falls perfectly in LISA's most sensitive frequency band, any deviation from a power-law behavior in the unperturbed spectrum can produce spectral features that allow the inference to break the degeneracy between the newly introduced diffusion and drift parameters $\kappa_{1,2}$ and the ones used to model the unperturbed GW background, as the perturbations depend on the derivatives in frequency of the unperturbed spectrum. 

From figure \ref{fig:corner_plot} we find that the parameters $\kappa_{1,2}$ can be constrained at the 1$\sigma$ level to $\kappa_1\leq 1.8\times 10^{-56}\text{kg}\,\text{m}^2\,\text{s}^{-3}$ and $\kappa_2\leq 3.7\times 10^{-56}\text{kg}\,\text{m}^2\,\text{s}^{-3}$ under the detection of a first-order phase transition GW background, and to $\kappa_1\leq 6.5\times 10^{-60}\text{kg}\,\text{m}^2\,\text{s}^{-3}$ and $\kappa_2\leq 2.6\times 10^{-59}\text{kg}\,\text{m}^2\,\text{s}^{-3}$ in the PBH scenario.
It is important to note that the first-order phase transitions and PBH backgrounds considered in \cite{Caprini:2024hue} and \cite{Bartolo:2018rku}, respectively, have relatively high amplitudes, while other scenarios may produce a possibly smaller signal.
An inspection of the Fisher matrix reveals the rough dependence of the 1$\sigma$ constraint on $\kappa_{1,2}$ in terms of $\Omega_b$, provided that the peak of the signal stands in the LISA sensitivity band and is not excessively broad. This is given by 
\begin{align}
\label{kappa_sigma_forecast}
\sigma_{\kappa_1}\simeq\sigma_{\kappa_2}\simeq\left(\frac{10^{-11}}{\Omega_b} \right)10^{-56}\text{kg}\,\text{m}^2\,\text{s}^{-3}\,.
\end{align}

%
%

\section{Conclusion}
\label{conclusion_sec}

Assuming Lorentz invariance to be an exact symmetry of quantum gravity that is also preserved effectively at low energies, we analyzed the covariant diffusion model of massless particles. The model has already been applied to photons and constrained via CMB blackbody spectrum \cite{philpott2009energy}. Here we solved the modified Boltzmann equation analytically at the background FRW level and showed explicitly that the effect dominates at low frequencies. This contrasts with the intuitive expectation that high-energy particles should be more sensitive to new Planck-scale physics. Nevertheless, this low-frequency dominance arises as a consequence of Lorentz invariance. As a result, quantum gravity phenomenology in this case would manifest at the largest scales.

We then performed a Fisher forecast on the two parameters of the model for LISA. Assuming a bright enough background signal either from first-order phase transitions or second-order gravitational waves from large scalar perturbations, we found that we will be able to put stringent bounds on the diffusion and drift parameters of the model. This opens the possibility of detecting the covariant diffusion of gravitons with diffusion and drift constants beyond the sensitivity of CMB experiments. Note, however, that the diffusion and drift parameters might as well differ between species of massless particles, or even depend on Lorentz-invariant properties of a given waveform \cite{Dowker:2013dog}. It is therefore an interesting problem for future work to go beyond the point-particle or geometric optics limit and model the covariant diffusion of scalar or higher-spin waves; see \cite{Salcedo:2024nex} for some relevant progress. This will shed light on how $\kappa_{1,2}$ could depend on waveform properties and possibly the microscopic physics that is responsible for the diffusion. Ultimately, the interpretation of a bound like \eqref{kappa_sigma_forecast} will depend on the specific quantum gravity framework that gives rise to the diffusion.

It is important to remark that the present analysis with LISA assumes that all the other sources of resolved and unresolved GW signals are perfectly fitted in the datastream, the GW background is stationary and isotropic, and that the instrumental noise is Gaussian with a well-characterized power spectral density.
Even if we expect that some complications will be present in the LISA data analysis (along with several other experimental challenges), we would like to insist on the fact that the effect of graviton diffusion on the GW spectrum can be easily taken into account, as it depends on just the first two derivatives in frequency of the unperturbed graviton distribution function and introduces just two phenomenological parameters. Any GW background with a nontrivial frequency dependence is distorted by these diffusion effects in a very specific way, producing distinct signatures that are very difficult to mimic with some features of the unperturbed model.

\acknowledgments
We thank Fay Dowker, Carlo Contaldi, and Pulkit Ghoderao for many useful discussions and comments. This work was funded in part by STFC grant ST/W006537/1. G.M. acknowledges support from the Imperial College London Schr\"odinger Scholarship scheme. A.N. is
funded by the President’s PhD Scholarship from Imperial College London

\appendix

\section{Analytic solution}
\label{app:analytic_sol}
Starting from equation \eqref{eq:eqdiff_etaq}, we define the comoving energy as $\epsilon=aE$. It combines the two terms on the left into one single derivative.
Defining a new time parameter $dT=a^2d\eta$ further simplifies the left-hand side so that we get
\begin{equation}
\label{eq_f_conservation_explicit}
    \frac{\partial f}{\partial T}=\big(3\kappa_1-\kappa_2\big)\frac{1}{\epsilon^2}\frac{\partial}{\partial \epsilon}\big(\epsilon^2f\big)+\kappa_1\frac{1}{\epsilon^2}\frac{\partial}{\partial \epsilon}\Big(\epsilon^3\frac{\partial}{\partial \epsilon}f\Big)\,.
\end{equation}
Now we define a new function
\begin{equation}
    g(\epsilon,T)=\epsilon^{3-\frac{\kappa_2}{2\kappa_1}}f(\epsilon,T)\,.
\end{equation}
This removes the first-order derivatives in $\epsilon$:
\begin{equation}
    \frac{\partial g}{\partial T}=\kappa_1\epsilon\partial_{\epsilon}^2g+\kappa_3 g/\epsilon\,,
\end{equation}
where
\begin{equation}
    \kappa_3=\frac{\kappa_2}{4\kappa_1}(2\kappa_1-\kappa_2)\,.
\end{equation}
One can see that the above PDE for $g$ is scale-invariant with respect to the following transformations
\begin{equation}
    g\rightarrow\xi^c g\,,\quad T\rightarrow\xi^a T\,,\quad \epsilon\rightarrow\xi^a \epsilon\,.
\end{equation}
A similarity solution is one that is invariant with respect to this re-scaling (note that the PDE being symmetric does not imply that all of its solutions must be symmetric as well). Therefore, the class of symmetric solutions is as follows
\begin{equation}
    g(\epsilon,T)=T^d\mathcal{G}(\epsilon/T)\,,
\end{equation}
where $d=c/a$ is an arbitrary real number. $\mathcal{G}$ satisfies an ODE with variable $y=\epsilon/T$:
\begin{equation}
    \kappa_1\mathcal{G}''+\mathcal{G}'+(\frac{\kappa_3}{y^2}-\frac{d}{y})\mathcal{G}(y)=0\,.
\end{equation}
Once again, we can remove the first derivative by a redefinition
\begin{equation}
    G(y)=e^{y/(2\kappa_1)}\mathcal{G}\,,
\end{equation}
to get
\begin{equation}
    \kappa_1G''+(\frac{\kappa_3}{y^2}-\frac{d}{y}-\frac{1}{4\kappa_1})G=0\,.
\end{equation}
This equation can be mapped to a radial Schrodinger equation with Coulomb potential if we make the following identification
\begin{align}
\label{ll+1}
    &\ell(\ell+1)=-\frac{\kappa_3}{\kappa_1}\,,\\
    &2me^2/\hbar^2=-d/\kappa_1\,,\\
    &2m\mathcal{E}/\hbar^2=-\frac{1}{4\kappa_1^2}\,.
\end{align}
The last equation informs that the energy level of the equivalent hydrogen atom has to be negative, $\mathcal{E}<0$, so we should indeed use the bound states. The first equation implies that $\ell$ needs not to be an integer in this matching. The bound-state solution of the equivalent problem is \cite{Sakurai2011zz}
\begin{equation}
    G(y)=\Big(\frac{y}{2\kappa_1}\Big)^{\ell+1}e^{-y/(2\kappa_1)}F(\ell+1+d;2(\ell+1);y/\kappa_1)\,,
\end{equation}
where $F$ is the confluent hypergeometric function. We have to select an orthonormal subset of these functions by suitably choosing $d$. One way is to enforce
\begin{equation}
    d=-(\ell+1)-n\,,
\end{equation}
where $n$ is a natural number. Then $F$ will reduce to a generalized Laguerre polynomial of degree n. 
\begin{equation}
    F\Big(-n;1+(2\ell+1);\frac{y}{\kappa_1}\Big)=(-1)^nn!L_n^{(2\ell+1)}\Big(\frac{y}{\kappa_1}\Big)\,.
\end{equation}
The orthogonality relation of these polynomials is known. Therefore, we have found a complete basis of solutions to our diffusion equation, which we can use to expand any solution with similar limiting behavior as $\epsilon\rightarrow0$:
\begin{equation}
\label{Laguerre_expansion}
    f(T,\epsilon)=e^{-\frac{\epsilon}{\kappa_1T}}\frac{\epsilon^{\frac{\kappa_2}{2\kappa_1}-3}}{T^{\ell+1}}\Big(\frac{\epsilon}{2\kappa_1T}\Big)^{\ell+1}\sum_{n=0}^{\infty}(-1)^nn!c_n\frac{1}{T^n}L_n^{(2\ell+1)}\Big(\frac{\epsilon}{\kappa_1T}\Big)\,.
\end{equation}
The coefficients $c_n$ depend on the initial condition at $T=T_i$ and can be found using the orthogonality of Laguerre polynomials. The result is
\begin{equation}
\label{c_n_eq}
    c_n=\Big(\frac{2}{\kappa_1}\Big)^{\ell+1}(-1)^n\frac{T_i^n}{\Gamma(n+2\ell+2)}\int_{0}^{\infty}d\epsilon\ f(T_i,\epsilon)\epsilon^{\ell+3-\frac{\kappa_2}{2\kappa_1}}L_n^{(2\ell+1)}\Big(\frac{\epsilon}{\kappa_1T_i}\Big)\,.
\end{equation}
This series expansion of the solution is of little use for initial conditions that are not in the form of a simple exponential function times a rational function of $\epsilon$. In particular, it poorly approximates Gaussian functions in $\epsilon$, requiring many terms and highly accurate integration to determine the coefficients~$c_n$. Therefore, the above solution is of little practical use. However, it can be used to construct heat kernels for the diffusion equation. These heat kernels, in turn, can easily give us the general solution as an integral over energy rather than a series expansion.

We define the heat kernel of the massless diffusion equation to be $K(T,\epsilon;\epsilon_i)$ such that it satisfies
\begin{align}
\label{app_Kernel_eq}
    \frac{\partial}{\partial T}K=\big(3\kappa_1-\kappa_2\big)\frac{1}{\epsilon^2}\frac{\partial}{\partial \epsilon}\big(\epsilon^2K\big)+\kappa_1\frac{1}{\epsilon^2}\frac{\partial}{\partial \epsilon}\Big(\epsilon^3\frac{\partial}{\partial \epsilon}K\Big)\,,
\end{align}
with the initial condition
\begin{align}
    K(T_i,\epsilon;\epsilon_i)=\delta(\epsilon-\epsilon_i)\,.
\end{align}
Substituting this delta function instead of $f(T_i,\epsilon)$ in \eqref{c_n_eq} gives the expansion coefficients $c_n[K]$ of~$K$
\begin{align}
    c_n[K]=\Big(\frac{2}{\kappa_1}\Big)^{\ell+1}\frac{(-1)^{n}}{\Gamma(n+2\ell+2)}T_i^n\epsilon_i^{\ell+3-\frac{\kappa_2}{2\kappa_1}}L_n^{(2\ell+1)}\Big(\frac{\epsilon_i}{\kappa_1T_i}\Big)\,.
\end{align}
Plugging this expression for the coefficients back into the expansion \eqref{Laguerre_expansion} for $K$ gives
\begin{equation}
\label{Laguerre_expansion_h}
    K(T,\epsilon;\epsilon_i)=e^{-\frac{\epsilon}{\kappa_1T}}\frac{1}{\epsilon_i}\Big(\frac{\epsilon_i}{\epsilon}\Big)^{^{3-\frac{\kappa_2}{2\kappa_1}}}\Big(\frac{\epsilon\epsilon_i}{\kappa_1^2T^2}\Big)^{\ell+1}\sum_{n=0}^{\infty}\frac{n!}{\Gamma(n+2\ell+2)}\Big(\frac{T_i}{T}\Big)^nL_n^{(2\ell+1)}\Big(\frac{\epsilon}{\kappa_1T}\Big)L_n^{(2\ell+1)}\Big(\frac{\epsilon_i}{\kappa_1T_i}\Big)\,.
\end{equation}
This cumbersome expression would not be useful unless we have an analytic expression for the series. Surprisingly enough, this exists and is given by the Hardy-Hille formula:
\begin{align}
    \sum_{n=0}^{\infty}\frac{n!}{\Gamma(n+1+\alpha)}t^nL_n^{(\alpha)}(x)L_n^{(\alpha)}(y)=\frac{1}{(xyt)^{\alpha/2}(1-t)}e^{-\frac{(x+y)t}{1-t}}I_\alpha\Big(\frac{2\sqrt{xyt}}{1-t}\Big)\,,
\end{align}
where $I_\alpha$ is the modified Bessel function. Therefore, we now have a closed-form expression for the heat kernel $K$. However, note that since equation \eqref{ll+1} has two solutions
\begin{equation}
    2\ell_\pm+1=\pm(1-\frac{\kappa_2}{\kappa_1})\,,
\end{equation}
we have actually found two corresponding heat kernels $K_\pm$ given in equation \eqref{h_closed_form}. Now we demand the total particle number to be finite and conserved. In terms of the heat kernels, we demand $\int_0^\infty d\epsilon\ \epsilon^2K_\pm$ to be equal to $\epsilon_i^2$ at all times. To see whether either of $K_\pm$ satisfies this condition, we should perform a complicated weighted integral over the modified Bessel function. A simpler way is to use the PDE governing $K_\pm$. Multiplying equation \eqref{app_Kernel_eq} by $\epsilon^2$ and integrating gives total derivative integrals on the right. So we only get contributions from $\epsilon=0,\infty.$ One can see that at $\epsilon\rightarrow\infty$, both $K_\pm$ go to zero. Hence
\begin{equation}
\label{partial_T_n}  \partial_T\int_0^\infty d\epsilon\ \epsilon^2K_\pm=-(3\kappa_1-\kappa_2)\epsilon^2K_\pm\Big|_{\epsilon\rightarrow0}-\kappa_1\epsilon^3\partial_\epsilon K_\pm\Big|_{\epsilon\rightarrow0}\,.
\end{equation}
Therefore, we would have to know the limiting behavior of $K_\pm$ as $\epsilon\rightarrow0$. Using the limit of the modified Bessel function, one can show that for $\epsilon\ll \kappa_1\Delta T$,
\[
K_+ \sim
\left\{
\begin{aligned}
  \frac{1}{\Gamma(2-\kappa_2/\kappa_1)}\left(\frac{\epsilon_i}{\kappa_1\Delta T}\right)^{2-\kappa_2/\kappa_1}e^{-\frac{\epsilon+\epsilon_i}{\kappa_1\Delta T}}\epsilon_i/\epsilon^2\,,\ \ & \text{if }\kappa_2/\kappa_1\notin\{1,2,3,...\} \\
  \frac{1}{\Gamma(\kappa_2/\kappa_1)}\left(\frac{1}{\kappa_1\Delta T}\right)^{\kappa_2/\kappa_1}e^{-\frac{\epsilon+\epsilon_i}{\kappa_1\Delta T}}\epsilon_i^2\epsilon^{\kappa_2/\kappa_1-3}\,,\ \ & \text{if }\kappa_2/\kappa_1\in\{1,2,3,...\}
\end{aligned}
\right.\,,
\]
\[
K_- \sim
\left\{
\begin{aligned}
  \frac{1}{\Gamma(2-\kappa_2/\kappa_1)}\left(\frac{\epsilon_i}{\kappa_1\Delta T}\right)^{2-\kappa_2/\kappa_1}e^{-\frac{\epsilon+\epsilon_i}{\kappa_1\Delta T}}\epsilon_i/\epsilon^2\,,\ \ & \text{if }\kappa_2/\kappa_1\in\{0,-1,-2,...\} \\
  \frac{1}{\Gamma(\kappa_2/\kappa_1)}\left(\frac{1}{\kappa_1\Delta T}\right)^{\kappa_2/\kappa_1}e^{-\frac{\epsilon+\epsilon_i}{\kappa_1\Delta T}}\epsilon_i^2\epsilon^{\kappa_2/\kappa_1-3}\,,\ \ & \text{if }\kappa_2/\kappa_1\notin\{0,-1,-2,...\}
\end{aligned}
\right.\,.
\]
Using these limit forms in \eqref{partial_T_n}, we find simple ODEs for the particle number as a function of time, which can easily be solved and give the results in the main text \eqref{particle_number_for_heat_kernels} for the total particle number.

\section{Fisher analysis}\label{app:Fisher}

The Fisher analysis involves performing the Taylor expansion of the exponent in the likelihood from equation \eqref{eq:Likelihood} as a function of the theoretical parameters $\theta$ around its minimum\footnote{For a detailed derivation of the Fisher analysis for cosmological observables, we refer to \cite{2020_cosmo_book}}. The parameter set that minimizes the chi-squared is given by the fiducial values assumed in the analysis, and is denoted by $\bar\theta$. The gradient of the chi-squared at $\bar\theta$ vanishes, and therefore the first non-vanishing term of the expansion comes from its Hessian, which is called the Fisher matrix, evaluated at the fiducial values
\begin{align}
\chi^2(\theta)&\simeq
(\theta-\bar\theta)^\alpha\cdot\mathcal{F}_{\alpha\beta}(\bar\theta)\cdot(\theta-\bar\theta)^\beta\,,\nonumber\\
\mathcal{F}_{\alpha\beta}&=\frac{1}{2}\frac{\partial^2\chi^2(\theta)}{\partial\theta^\alpha\partial\theta^\beta}_{\vert_{\theta=\bar\theta}}
=T_{\rm obs}\int_{\nu_{\rm min}}^{\nu_{\rm max}} d\nu\,\frac{\partial \Omega_{\rm GW}(\nu;\theta)}{\partial\theta^\alpha}_{\vert_{\theta=\bar\theta}}\frac{\partial \Omega_{\rm GW}(\nu;\theta)}{\partial\theta^\beta}_{\vert_{\theta=\bar\theta}}\,\frac{4\pi}{\Omega_{n}^2(\nu)}\,,
\end{align}
where the last equality holds since we are expanding the chi-squared around its minimum.
In our model, we identify the set of parameters of the unperturbed model as $\theta_0$, and the two newly introduced diffusion and drift parameters are $\kappa_{1,2}$.
The explicit evaluation of the Fisher matrix simplifies in the limit where $\kappa_{1,2}\ll 1$:
\begin{align}
\mathcal{F}_{\theta_{0\,a},\theta_{0\,b}}&=T_{\rm obs}\int_{\nu_{\rm min}}^{\nu_{\rm max}} d\nu\,\frac{\partial \Omega_0(\nu;\theta_{0})}{\partial\theta_{0\,a}}_{\vert_{\theta_0=\bar\theta_0}}\frac{\partial \Omega_0(\nu;\theta_0)}{\partial\theta_{0\,b}}_{\vert_{\theta_0=\bar\theta_0}}\,\frac{4\pi}{\Omega_{n}^2(\nu)}\,,\nonumber\\
\mathcal{F}_{\theta_{0\,a},\kappa_i}&=T_{\rm obs}\int_{\nu_{\rm min}}^{\nu_{\rm max}} d\nu\,\frac{\partial \Omega_0(\nu;\theta_0)}{\partial\theta_{0\,a}}_{\vert_{\theta_0=\bar\theta_0}}\Omega_i(\nu;\theta_0)_{\vert_{\theta_0=\bar\theta_0}}\,\frac{4\pi}{\Omega_{n}^2(\nu)}\,,\nonumber\\
\mathcal{F}_{\kappa_i,\kappa_j}&=T_{\rm obs}\int_{\nu_{\rm min}}^{\nu_{\rm max}} d\nu\,\Omega_i(\nu;\theta_0)_{\vert_{\theta_0=\bar\theta_0}}\Omega_j(\nu;\theta_0)_{\vert_{\theta_0=\bar\theta_0}}\,\frac{4\pi}{\Omega_{n}^2(\nu)}\,.\nonumber\\
\end{align}
The forecast likelihood function of the theoretical parameters can be approximated by a multivariate normal distribution with $\bar\theta$ as its mean and the inverse of the Fisher matrix as its covariance matrix:
\begin{align}
\mathcal{L}(\theta)&\simeq \mathcal{N}(\bar\theta,\Sigma^F)\,,\nonumber\\
\Sigma^F&=\mathcal{F}^{-1}\,.
\end{align}
For the case of the GW background considered in \ref{subsec:PT_PBH}, the Fisher matrix is
\begin{align}
\mathcal{F}&=\begin{pmatrix}
\mathcal{F}_{\Omega_b,\Omega_b} & \mathcal{F}_{\Omega_b,\nu_b} & \mathcal{F}_{\Omega_b,\kappa_1} & \mathcal{F}_{\Omega_b,\kappa_2} \\
\mathcal{F}_{\nu_b,\Omega_b} & \mathcal{F}_{\nu_b,\nu_b} & \mathcal{F}_{\nu_b,\kappa_1} & \mathcal{F}_{\nu_b,\kappa_2} \\
\mathcal{F}_{\kappa_1,\Omega_b} & \mathcal{F}_{\kappa_1,\nu_b} & \mathcal{F}_{\kappa_1,\kappa_1} & \mathcal{F}_{\kappa_1,\kappa_2} \\
\mathcal{F}_{\kappa_2,\Omega_b} & \mathcal{F}_{\kappa_2,\nu_b} & \mathcal{F}_{\kappa_2,\kappa_1} & \mathcal{F}_{\kappa_2,\kappa_2}
\end{pmatrix}\,,\nonumber\\
\mathcal{F}_{\Omega_b,\Omega_b}&=T_{\rm obs}\int_{\nu_{\rm min}}^{\nu_{\rm max}} d\nu\,\psi_0^2(\nu;\bar \nu_b)\,\frac{4\pi}{\Omega_{n}^2(\nu)}\,,\nonumber\\
\mathcal{F}_{\Omega_b,\nu_b}&=\bar\Omega_b T_{\rm obs}\int_{\nu_{\rm min}}^{\nu_{\rm max}} d\nu\,\psi_0(\nu;\bar \nu_b)\,\left(\frac{\partial \psi_0(\nu;\nu_b)}{\partial \nu_b}_{\vert_{\nu_b=\bar \nu_b}}\right)\,\frac{4\pi}{\Omega_{n}^2(\nu)}\,,\nonumber\\
\mathcal{F}_{\nu_b,\nu_b}&=\bar\Omega_b^2 T_{\rm obs}\int_{\nu_{\rm min}}^{\nu_{\rm max}} d\nu\,\left(\frac{\partial \psi_0(\nu;\nu_b)}{\partial \nu_b}_{\vert_{\nu_b=\bar \nu_b}}\right)^2\,\frac{4\pi}{\Omega_{n}^2(\nu)}\,,\nonumber\\
\mathcal{F}_{\kappa_{i},\nu_b}&=\bar\Omega_b^2 T_{\rm obs}\int_{\nu_{\rm min}}^{\nu_{\rm max}} d\nu\,\psi_{i}(\nu;\bar \nu_b)\,\left(\frac{\partial \psi_0(\nu;\nu_b)}{\partial \nu_b}_{\vert_{\nu_b=\bar \nu_b}}\right)\,\frac{4\pi}{\Omega_{n}^2(\nu)}\,,\nonumber\\
\mathcal{F}_{\kappa_{i},\kappa_{j}}&=\bar\Omega_b^2 T_{\rm obs}\int_{\nu_{\rm min}}^{\nu_{\rm max}} d\nu\,\psi_{i}(\nu;\bar \nu_b)\,\psi_{j}(\nu;\bar \nu_b)\,\frac{4\pi}{\Omega_{n}^2(\nu)}\,,\nonumber\\
\mathcal{F}_{\Omega_b,\kappa_i}&=\bar\Omega_b T_{\rm obs}\int_{\nu_{\rm min}}^{\nu_{\rm max}} d\nu\,\psi_0(\nu;\bar \nu_b)\,\psi_i(\nu;\bar \nu_b)\,\frac{4\pi}{\Omega_{n}^2(\nu)}\,,
\end{align}
with $\bar\theta=\{\bar\Omega_b,\bar \nu_b,\bar\kappa_1,\bar\kappa_2\}$ being the fiducial parameters. 
The forecast error on $\kappa_1$ or $\kappa_2$ after marginalizing the posterior distribution over the other parameters is the square root of the $(\kappa_1,\kappa_1)$ or $(\kappa_2,\kappa_2)$ element of the inverse Fisher matrix, i.e.
\begin{align}
\sigma_{\kappa_1}^{\rm marginal}=\sqrt{\Sigma^F_{\kappa_1\kappa_1}}\,,\nonumber\\
\sigma_{\kappa_2}^{\rm marginal}=\sqrt{\Sigma^F_{\kappa_2\kappa_2}}\,.
\end{align}

\bibliographystyle{JHEP}
\bibliography{main.bib}

\end{document}